\documentclass[journal=jacsat,manuscript=article]{achemso}

\usepackage{chemformula} 
\usepackage[T1]{fontenc} 
\usepackage{multirow}
\usepackage{graphicx}
\usepackage{caption}

\author{Xiangrui Yang}
\email{yangxr9@mail3.sysu.edu.cn}

\title
  {Leveraging Orbital Information and Atomic Feature in Deep Learning Model }

\abbreviations{IR,NMR,UV}

\begin{document}







\begin{abstract}

Predicting material properties base on micro structure of materials has long been a challenging problem. Recently many deep learning methods have been developed for material property prediction. In this study, we propose a crystal representation learning framework, Orbital CrystalNet, OCrystalNet, which consists of two parts: atomic descriptor generation and graph representation learning. In OCrystalNet, we first incorporate orbital field matrix (OFM) and atomic features to construct OFM-feature atomic descriptor, and then the atomic descriptor is used as atom embedding in the atom-bond message passing module which takes advantage of the topological structure of crystal graphs to learn crystal representation. To demonstrate the capabilities of OCrystalNet we performed a number of prediction tasks on Material Project dataset and JARVIS dataset and compared our model with other baselines and state of art methods. To further present the effectiveness of OCrystalNet, we conducted ablation study and case study of our model. The results show that our model have various advantages over other state of art models.

\end{abstract}

\section{Introduction}

Predicting material properties has long been a challenging subject which is vital in developing new materials. Over the years, scientists have been using laboratory experiments or computational quantum mechanics modeling, namely Density Functional Theory (DFT) to investigate material properties\cite{ramprasad2017machine, parr1983density}. Though, laboratory experiments have the advantage of high accuracy and reliability, this method costs researchers’ large amount of time and effort \cite{parr1983density}. Worse still it is almost impossible to conduct laboratory experiments on large dataset. With the help of high performance clusters, researchers can run DFT on large dataset and get relatively reliable results on thousands of materials \cite{neugebauer2013density}. However, high performance clusters are relatively scarce under current research environment leading to the fact that DFT implementation is still uncommon right now \cite{cohen2012challenges}. With the fast developing of machine learning algorithms and deep learning models, increasing efforts are devoted to predicting material properties using these methods. Owning to the fast calculation speed and improving accuracy, machine learning algorithms are widely adopted to accelerate material development\cite{xie2018crystal, karamad2020orbital, yang2012search, ortiz2009data}. Roughly speaking, machine learning algorithms are functions which map vector representing materials to target properties. Hence, it is important to design a descriptor for materials which contains rich information about material chemistry properties and is compatible with machine learning algorithms \cite{mansimov2019molecular}. Another important step is to leverage the descriptor and make accurate prediction, i.e. design the mapping function. Therefore we are going to 
address both the challenges in following paragraphs.

First, the properties of crystalline materials are largely determined by the atom types and the spacial structure determined by these atoms. To capture information of atoms, a number of atomic descriptors have been proposed by researchers \cite{mansimov2019molecular}. Behler et al. proposed an atomic descriptor based on atom-distribution-based symmetry functions, which was later used to calculate total energy of materials \cite{behler2007generalized}. Another descriptor, Coulumb Matrix was made by Rupp et al. whose dimension equals to the total number of atoms\cite{rupp2012fast}. Xie et al. combined ont-hot embedding and a few features of atoms to generate a new descriptor which satisfies the requirement of atom uniqueness and reflects basic atomic features \cite{xie2018crystal}. A number of works followed or modified based on above descriptors and achieved relatively accurate result \cite{chen2019graph, schutt2018schnet}. Another way of describing atoms is dedicated to representing the electronic configuration of each atom. Isayev et al. proposed an electrical descriptor which incorporates band structures and density of states descriptor \cite{isayev2015materials}. Later, based on the thought of describing center atom with the help of surrounding atoms and its own electronic configuration in crystaline materials, Pham et al. developed orbital field matrix (OFM) to encode atomic valence orbital interactions \cite{pham2017machine}. Since OFM successfully capture the atomic electron configuration and valence orbital interactions which in turn determine the spacial structure of crystalline material, it is relatively efficient in deep learning models with a few modifications proposed by Karamad et al. \cite{karamad2020orbital} Karamad et al. enhanced the importance of central atom in the atomic descriptor and proposed orbital graph convolutional neural network (OGCNN) \cite{karamad2020orbital} to capture the information. Despite OFM and OGCNN exhibit high capability of incorporating atomic valence orbital information, they do not contain enough information of atomic features and more importantly they failed to fully leverage topological structure of crystal graphs to learn crystalline material representation.

There have been a lot of deep learning models developed for material property prediction which have achieved high accuracy \cite{karamad2020orbital, xie2018crystal, schutt2018schnet, chen2019graph}. Under the framework of Message Passing Neural Network (MPNN)\cite{gilmer2017neural}, Xie at el. proposed crystal graph convolutional neural network (CGCNN) which first views periodic crystalline material in graph manner \cite{xie2018crystal}. The model takes in one-hot embedding and a few basic atomic features to calculate the overall vector representation of crystalline materials using plain graph convolutional neural network. CGCNN ignores the message passed from nodes to edges which leads to the failure of fully utilizing atomic and edge information \cite{xie2018crystal}. Based on previous work\cite{schutt2017quantum}, Schutt et al. constructed SchNet containing continuous-filter convolutional layers which implicitly passes the edge and neighbor message to nodes and add atom-wise layers taking advantage of nodes' own features \cite{schutt2018schnet}. Also under the framework of MPNN, Chen et al. proposed Material Graph Network (MEGNet) which introduced global state attributes into the process of message passing \cite{chen2019graph}. Although, MEGNet leverages global state attributes to capture information of global environment of material, the model creates unnecessary loops in message passing trajectory and ignores the direction of edge message passing process. Incorporating the geometric information in micro structures, Cheng et al. proposed geometric-information-enhanced crystal graph neural network (GeoCGNN) which achieves relatively great results and proves insights on the influences of geometric information on material properties \cite{cheng2021geometric}. Most recently, Choudhary et al. proposed Atomistic Line Graph Neural Network (ALIGNN) which reconstructs the graph representation of crystal in order to encode the angles between bonds \cite{choudhary2021atomistic}. Despite ALIGNN exhibits superior performance over other models, it suffers from the complicated model structure which leads to a large increase of parameters the model. 

To handle the problems mentioned above, we combine the OFM and features of atoms together to construct OFM-feature descriptor. The OFM-feature descriptors use large amount of atomic features to distinguish every atom in hundreds of dimensions and adds the valence orbital information as a important factor in determining material properties. As for the construction of deep learning model, we present a novel model named Orbital CrystalNet. Inspired by recent work Communicative Message Passing Neural Network (CMPNN)\cite{song2020communicative}, which improves molecular embedding by strengthening message passing between molecular atoms and bonds, OCrystalNet considers the interaction between bonds and atoms in crystalline materials. At the same time OCrystalNet is able to maintain size invariance and permutational invariance of crystal representation.


To illustrate the performance of our proposed model, we conducted material predicting experiments on Material Project \cite{jain2013commentary} and newly proposed JARVIS dataset \cite{choudhary2020joint}. We also conducted ablation study to prove effectiveness of atomic descriptor. To illustrate the correctness of our predicting results, we conducted case study of several materials. Finally, we provide a detailed analysis of OCrystalNet with experiments of variable training size and prediction results against DFT calculated results. 

In summary, we provide the following main contributions in this study:

(1) We propose OFM-feature descriptor as atomic descriptor in deep learning model which incorporates both valence orbital information and atomic features.

(2) We propose a deep learning model named Orbital CrystalNet with strong ability in capturing the atomic information and topological structure of crystal graphs.

(3) We conduct extensive experiments on different public datasets to prove the effectiveness of our model and different modules in our model.

\section{Method}

\subsection{Preliminary}

Before describing our model in full detail, we give a clear definition of all the concepts we use in the model. Then, we present our prediction task in a formal way.

In order to get a graph representation of a crystal structure, we need to statistically represent atoms and bonds in micro structures. In every crystal each atom $u$ is represented by a feature vector $x(u)$, the generation of which will be discussed in detail later in this section. Due to the periodicity of crystals, we allow multiple edges between a same pair of nodes. Thus, we denote a specific edge between node $u$, $v$ with $(u, v)_k$, and the feature vector of which is $x((u, v)_k)$. Finally we denote the crystal graph using $G = (E, V)$, in which $E$ and $V$ denote the set of nodes and edges respectively.

Our task of getting crystal graph representation is getting a function mapping the crystal graph $G$ to specific vector $x(G)$. As for down stream prediction tasks such as predicting the band gap and formation energy we need to obtain another mapping function to map the vector representation of crystal graph to target properties $\hat{y}$.

\subsection{Atomic Orbital Embedding}
The key idea of OCrystalNet is leveraging both the information of atoms in crystal and the topological structure. To construct atomic descriptor, we combine both orbital interactions between atoms and atomic features. 

\begin{figure}[h]
    \includegraphics[width=1\textwidth]{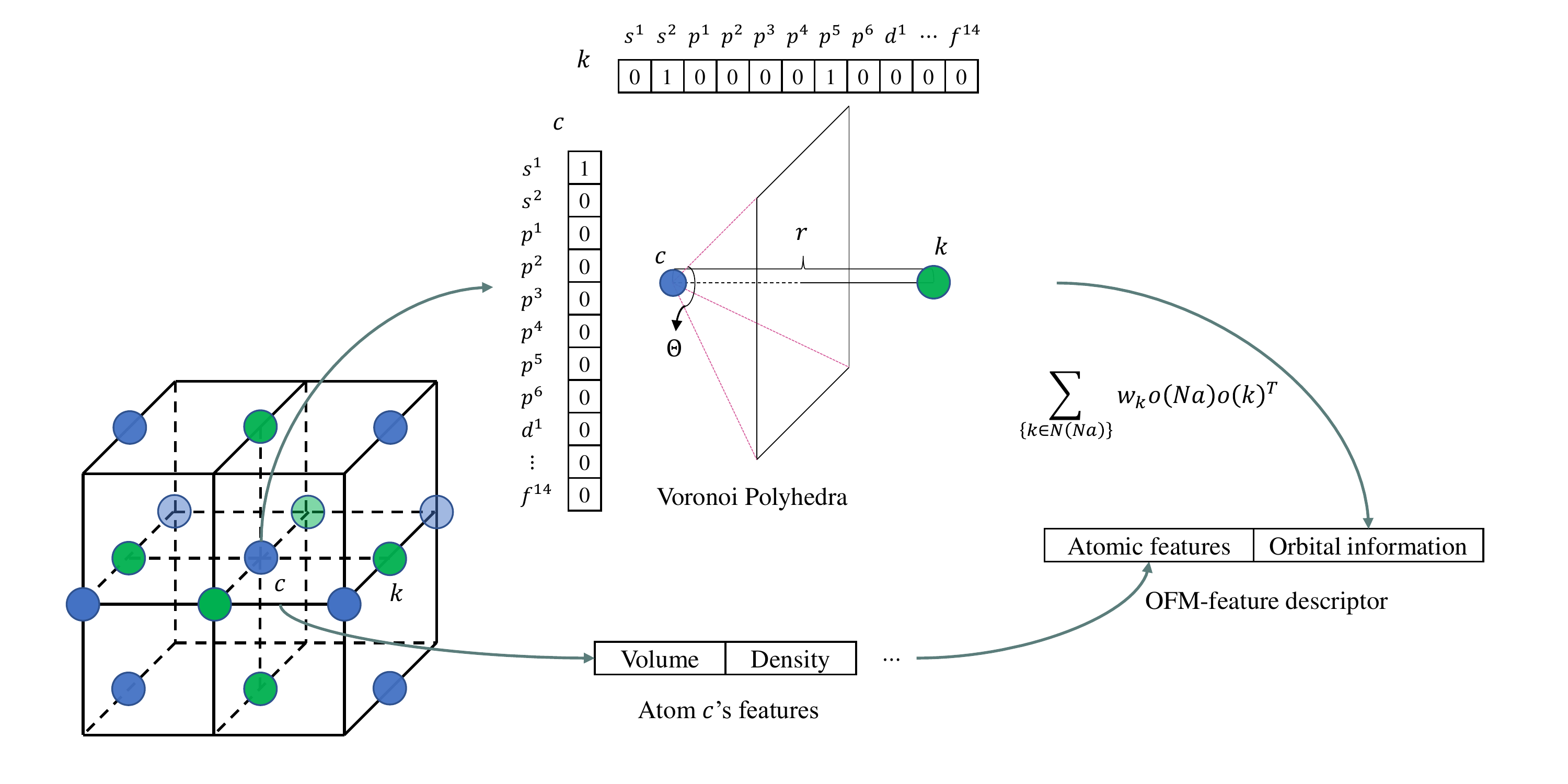}
    \caption{OFM-feature descriptor for atom embedding}
    \label{fg:atom_embedding}
\end{figure}

First, to incorporate the orbital valence information, we follow the work of Pham et al., encoding this information in one-hot vector $o(c) \in \mathcal{R}^{32}$.\cite{pham2017machine} The detailed setting of each dimension is shown in Figure \ref{fg:atom_embedding}. We use $c$ and $k$ to denote the center atom and surrounding atoms. After getting vector representation of electrical configuration of each atom, we use the surrounding atoms to calculate the Orbital Field Matrix (OFM) of the center atom using following equation:

\begin{equation}
    \label{eq:sum_ofm}
    X^{c} = \sum_{k \in N(c)} w_k o(c) o(k)^T
\end{equation}
where $N(c)$ denotes the neighbor set of atom $c$ and $w_k$ denotes the weight of each atom pair. The weight parameter is calculated based on the solid angle determined by the face of the Voronoi polyhedron $\theta_k$ and the distance of the atom pair $r_k$ as follow:

\begin{equation}
    \label{eq:ofm_weight}
    \begin{aligned}
    \theta_{max} &= max_{k \in N(c)} \{ \theta_k \} \\
    w_k &= \xi (r_k) \theta_k / \theta_{max}
    \end{aligned}
\end{equation}
where function $\xi(r_k) = \frac{1}{r_k}$ contains information of the strength of the bond between two atoms and valence orbitals of each atom. Figure \ref{fg:atom_embedding} demonstrates the construction process of Voronoi Polyhedra. Both solid angle $\theta$ and distance between atoms $r$ are marked in Figure \ref{fg:atom_embedding}. Later, we expand the 2D OFM by row and concatenate the original one-hot vector representing electron layer to it getting the OFM representation $x_o(c)$ of $1056$ dimensions.

Finally, we take advantage of atomic features to enrich the information in node embedding, after the process of compressing information of valence orbitals of atoms and the interactions between them. We obtain the all atomic features from JARVIS \cite{choudhary2020joint} which collects 438 properties of each atom/element such as the number of row in periodic table and the atom mass against $C^{12}$ etc. We denote the final atom embedding as follow:

\begin{equation}
    \label{eq:atom_embedding}
    x(c) = x_o(c) \oplus x_p(c)
\end{equation}
where $x_p(c)$ represents the physical properties of atoms/elements and $\oplus$ is the concatenate operation.

\subsection{Bond Embedding}

The bonds in crystal is not explicitly defined in chemistry. Therefore, the edges between nodes are defined if the distances between atom pairs $d(u, v)$ are under a specific cutoff parameter $d_0$. We follow the work of Xie et al. \cite{xie2018crystal}. to expand the distance to a vector $d((u, v)_k)$ and unify it with Gaussian filter as follow:

\begin{equation}
    \label{eq:bond_embedding}
    x((u, v)_k) = exp(-(d((u, v)_k) - \bar{d})^2 / \sigma^2)
\end{equation}
where $\bar{d} = 2.5$ and $\sigma = 0.5$.

\subsection{OCrystalNet Structure}

OCrystalNet is under the general frame work of Message Passing Neural Network (MPNN)\cite{gilmer2017neural} and is modified from Communicative Message Passing Neural Network (CMPNN)\cite{song2020communicative}.As general MPNN, OCrystalNet can be divided to two modules, namely message passing module and readout module. Figure 2 illustrates the overall structure of OCrystalNet.

\begin{figure}[h]
    \includegraphics[width=0.8\textwidth]{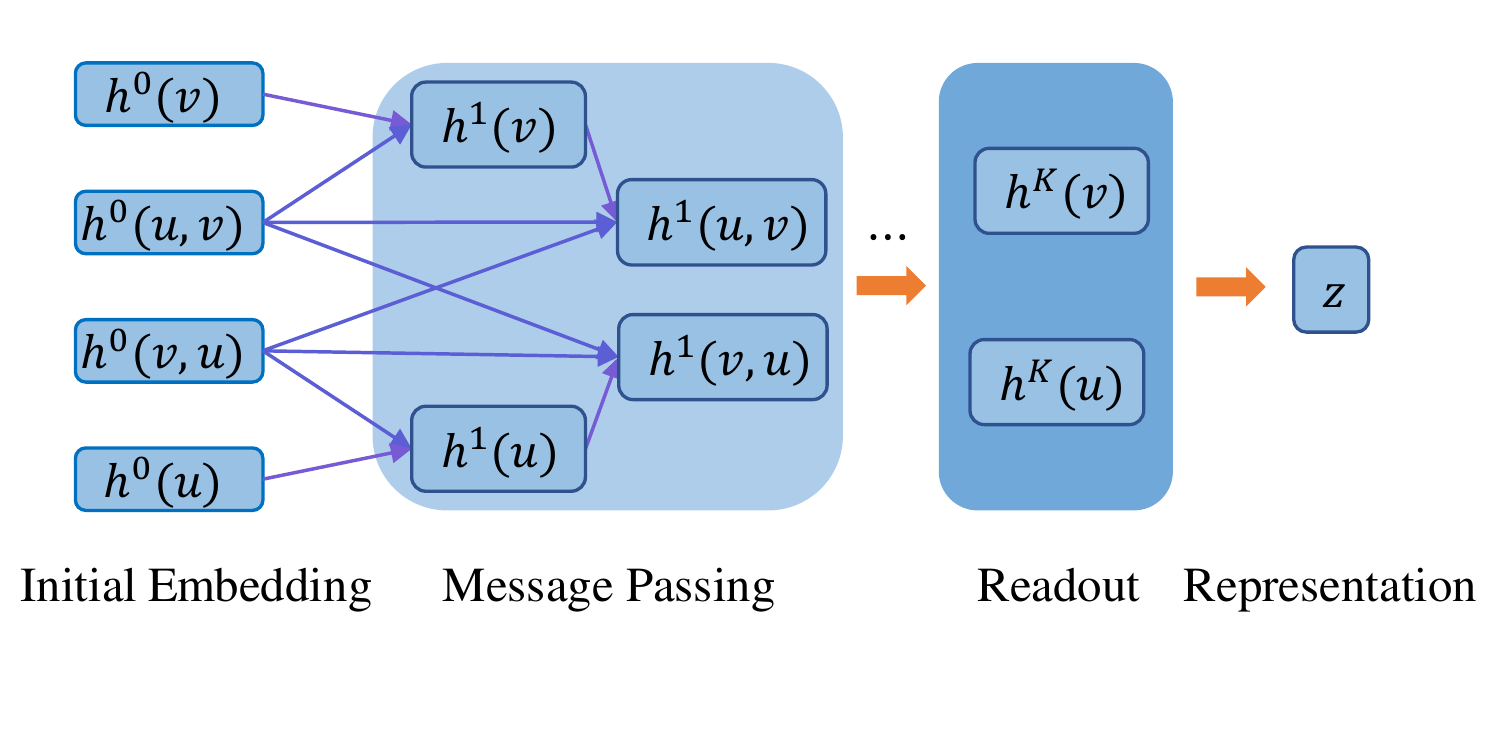}
    \caption{OCrystalNet Structure}
    \label{fg:OCrystalNet}
\end{figure}

In the message passing module, we implement standard massage passing layer which can be stacked $L$ times. Each layer leverages node hidden state $h(u)$, edge hidden state $h((u, v)_k)$, node message $m(u)$ and edge message $m((u, v)_k)$. For initial layer we pass the initial node embedding $x(u)$ and bond embedding $x((u, v)_k)$ as input hidden state. In layer $l \in \{1, 2, \cdots, L\}$ we first aggregate the incoming edge information to every node in the graph using following equation:

\begin{equation}
    \label{eq:aggregate_edge}
    m^l(u) = \sum_{v \in N(u)} \sum_k Aggregate(h^{l - 1}((v, u)_k))
\end{equation}
where $m^l(u)$ denotes the message vector of node $u$ in $l$ th layer and $h^{l - 1}((v, u)_k)$ denote the hidden state of directed edge $(v, u)_k$. The Aggregation function is set to be the same as CMPNN\cite{song2020communicative}. We need to point out that the direction of message aggregation is important, since the incoming message will later pass to outgoing edges and neighboring nodes. Next, we combine last layer node hidden state and the message vector to generate $l$ th layer node hidden state using a communicate function as follow:

\begin{equation}
    \label{eq:communicate_node}
    h^l(u) = Communicate(m^l(u), h^{l - 1}(u))
\end{equation}
where $h^l(u)$ denotes the hidden state of node $u$ in $l$ th layer. We propose our own Communicate function which achieves better performance on crystalline materials as follow:

\textbf{Linear Kernel.} Adding the two vector element-wise is time efficient and relatively consistent to the operation conducted in previous processing. Thus we implement the an element-wise add operation as follow:

\begin{equation}
    \label{eq:communicative_add}
    h^l(u) = m^l(u) + h^{l - 1}(u)
\end{equation}
where $+$ is element-wise add operation.



So far we finish processing node hidden state process and message passing, we then demonstrate how edge hidden state is handled. Similar to node processing, OCrystalNet first generate message vector for edges using the source node and its inverse edge information, which is formalized as follow:

\begin{equation}
    \label{eq:edge_message}
    m^l((u, v)_k) = h^l(u) - h^l((v, u)_k)
\end{equation}
where $m^l((u, v)_k)$ denotes the message vector of edge $(u, v)_k$ in $l$ th layer. Using source node hidden state subtracting inverse edge information avoids the problem of generating loops in message passing trajectory. Next, we generate the new hidden state of edge $(u, v)_k$ using message vector $m^l((u, v)_k)$ and initial hidden state as follow:

\begin{equation}
    \label{eq:edge_hidden}
    h^l((u, v)_k) = Mish(h^0((u, v)_k) + W^l m^l((u, v)_k))
\end{equation}
where $Mish(\cdot)$ is activation function and $W^l$ is a trainable parameter in layer $l$.

After $L$ layer of message passing, we will update node hidden follow the same pattern one more round to gather the information passing by last layer edge hidden state. Lastly, we implement readout function as mean pooling operation as follow:

\begin{equation}
    \label{eq:readout}
    z = \sum_{u \in V} h(u) / |V|
\end{equation}
where $|V|$ denote the size of node set $V$ and $z$ is the final graph representation generated by OCrystalNet. Bescause of the mean pooling operation is permutational invariance, the graph representation generated by OCrystalNet is not affected by the sequence of input atoms in crystal graphs.

\section{Experiment}

\subsection{Experiment setups}
\begin{figure}[htbp]
    \centering
    \begin{minipage}{0.5\linewidth}
		\centering
		\captionof{table}{Dataset}
        \begin{tabular}{lll}
        \hline
        Dataset                            & Property         & Size  \\ \hline
                                          & Band gap         & 69239 \\
                                          & Formation energy & 69239 \\
                                          & Bulk modulus     & 5830  \\
        \multirow{-4}{*}{Material Project} & Shear modulus    & 5830  \\ \hline
                                          & Band gap (OPT)        & 55712 \\
                                          & Band gap (MBJ)  & 18167\\
                                          & Formation energy & 55712 \\
                                          & Bulk modulus     & 18300 \\
        \multirow{-4}{*}{JARVIS}           & Shear modulus    & 18300\\ \hline
        \end{tabular}
        
        \label{tab:dataset}
    \end{minipage}
	\hfill
	\begin{minipage}{0.45\linewidth}
		\centering
		\vspace{3.5em}
		\includegraphics[width=1\linewidth]{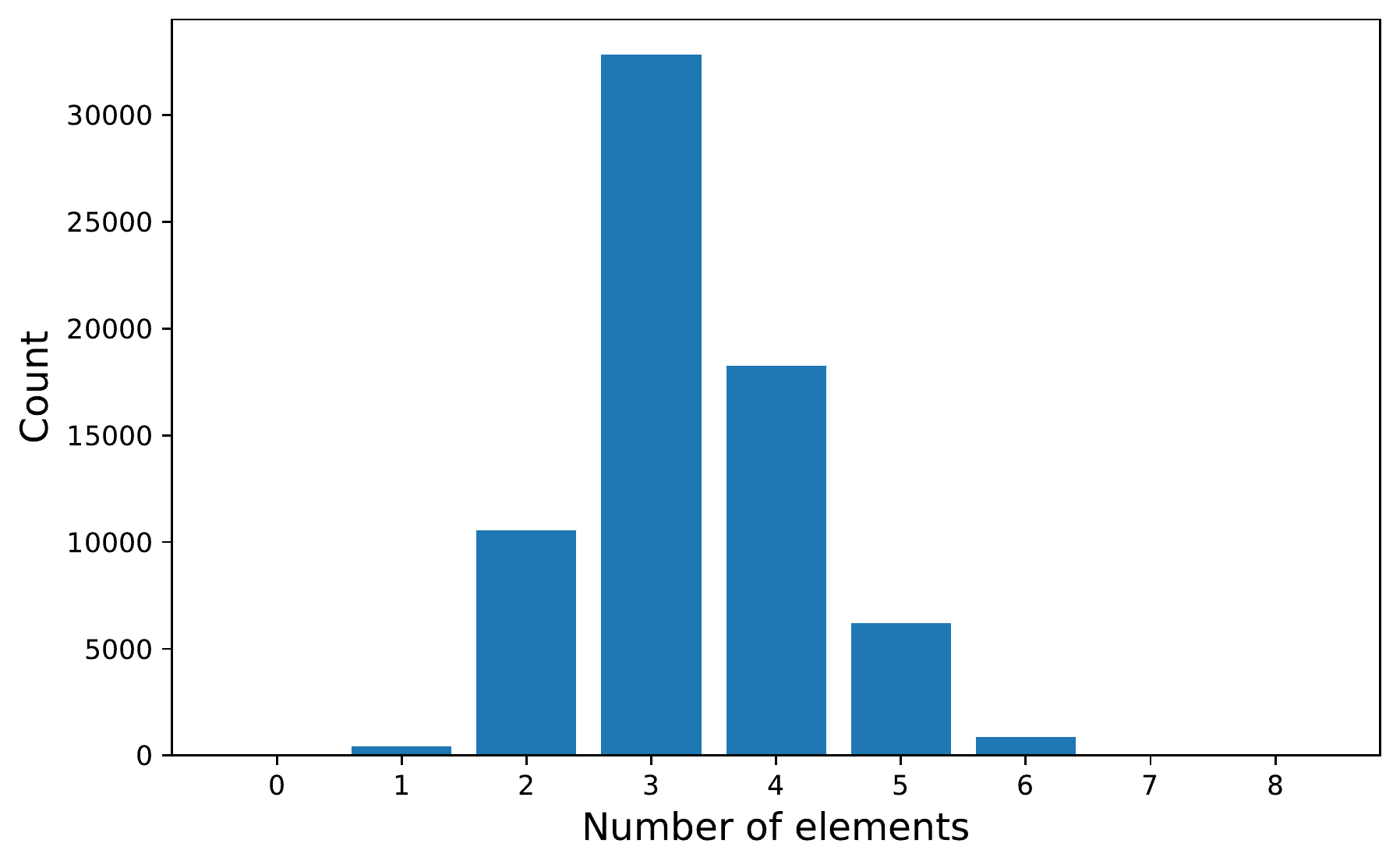}
		\caption{Crystal element distribution}
		\label{fg:dataset}
	\end{minipage}
\end{figure}

\textbf{Datasets.} In order to implement a comprehensive comparison with other benchmark algorithms, we evaluated our model in predicting four primary material properties, namely band gap $E_g$, formation energy $E_f$, bulk modulus $K_{VRH}$, and shear modulus $G_{VRH}$. The dataset are downloaded from Material Project \cite{jain2013commentary} and JARVIS \cite{choudhary2020joint}. The number of crystals in each datasets is shown in Table \ref{tab:dataset} and the distribution of crystal containing different number of elements are shown in Figure \ref{fg:dataset}.

\textbf{Baseline.} OCrystalNet was compared with 4 baseline deep learning methods in material property predicting all of which are listed as follow:

\begin{itemize}
    \item \textbf{CGCNN.} \cite{xie2018crystal} Xie et al. proposed the CGCNN based on MPNN and use a combination of one-hot embedding as atom descriptor.
    \item \textbf{SchNet.} \cite{schutt2018schnet} Schutt et al. improved their previous model\cite{schutt2017quantum} and proposed Schnet which includes a continuous-filter convolutional layer and leverages one-hot embedding as atom descriptor.
    \item \textbf{MAGNet.} \cite{chen2019graph} Chen et al. proposed MAGNet under the frame work of MPNN \cite{gilmer2017neural}. The model leverages global state variable and use one-hot embedding as tom descriptor.
    \item \textbf{OGCNN.} \cite{karamad2020orbital} Karamad et al. proposed OGCNN using a OFM as descriptor while the model is based on CGCNN.
    \item \textbf{GeoCGNN.} \cite{cheng2021geometric} Cheng et al. proposed GeoCGNN which leverage the geomertry information in crystal micro structures and leverages one-hot descriptor for atoms.
    \item \textbf{ALIGNN.} \cite{choudhary2021atomistic} Choudhary et al. proposed ALIGNN which takes in the angles between bonds as input and achieves state of art performance.
\end{itemize}

\textbf{Implementation Details.} For training models, we followed the set up of MEGNet\cite{chen2019graph}, using 60,000 crystals in Material Project \cite{jain2013commentary} for training and dividing the remaining crystals equally for validating (4619) and testing (4620). Later, We trained and tested four models on JARVIS to compare the generalization ability of each model. We split the dataset to 80:10:10 for training, validating and testing. Since all the prediction tasks are regression task, we reported Mean Absolute Error (MAE) of every prediction tasks. For OCrystalNet we set the best parameter for every prediction tasks and for other model we set the parameter as same as they mentioned in their articles. To achieve the full potential of our model, The Adam optimizer was used whose initial learning rate, max learning rate and final learning rate were set to 0.0001, 0.0003 and 0.00001 respectively. All models were trained and tested on Ubuntu Linux 16 with Nvidia GTX 1080Ti GPUs.

\subsection{Performance Comparison}
Table \ref{tab:Material Project} shows the performances evaluated by mean absolute error (MAE) of different models. Four tasks were conducted on Material Project \cite{jain2013commentary}, namely predicting band gap $E_g$, formation energy $E_f$, bulk modulus $K_{VRH}$ and shear modulus $G_{VRH}$. As shown in the result, our model, OCrystalNet, outperforms all other four models on band gap prediction by at least $21.12\%$. We contribute this superiority in predicting band gap to the rich information of valence orbital in atom embedding which is a essential factor of band gap. For the rest three property predictiong OCrystalNet reaches a improvment of $3.57\%$ on formation energy, $17.80\%$ on bulk modulus and $7.41\%$ on shear modulus, all of which is significally higher other comparing methods.

\begin{table}
  \caption{Comparison of MAEs of different models on Material Project}
  \label{tab:Material Project}
  \begin{tabular}{lllll}
    \hline
    Dataset   & Band Gap & Formation Energy & Bulk Modulus & Shear Modulus  \\
    \hline
    CGCNN       & 0.388 & 0.039  & 0.054 & 0.087 \\
    SchNet      & 0.362 & - & - & -\\
    MEGNet      & 0.330 $\pm$ 0.01 & 0.028 $\pm$ 0.000 & 0.050 $\pm$ 0.002 & 0.079 $\pm$ 0.003\\
    OGCNN       & 0.320 $\pm$ 0.01 & 0.030 $\pm$ 0.000 & 0.519 $\pm$ 0.002 & 0.081 $\pm$ 0.002\\
    GeoCGNN     & 0.281 $\pm$ 0.00 & 0.024 $\pm$ 0.000 & 0.057 $\pm$ 0.001 & 0.079 $\pm$ 0.000\\
    ALIGNN      & 0.218 & 0.022 & - & - \\
    OCrystalNet  & 0.252 $\pm$ 0.01 & 0.027 $\pm$ 0.000 & 0.041 $\pm$ 0.002& 0.073 $\pm$ 0.003\\
    \hline
    Improve     & 21.12\% & 3.57\% & 17.80\% & 7.41\%\\
    \hline
  \end{tabular}
\end{table}




To further illustrate the performance of OcrystalNet, we train our model on JARVIS and compare five prediction tasks with CGCNN and ALIGNN \cite{choudhary2020joint, choudhary2021atomistic, xie2018crystal}. The results are shown in Table \ref{tab:JARVIS}. As shown in the results, OCrystalNet performances well in band gap, bulk modulus and shear modulus predictions. We contribute the superiority in band gap prediction to the valence orbital information contained in atom embedding. As for bulk modulus and shear modulus, we contribute the superiority to atomic features encoded in atom embedding. It is worth noting that ALIGNN achieve state of art performance because it leverages the angle between bonds in crystals' micro structures. However, ALIGNN suffer from its complicated models structure which contains 4,026,753 parameters. In comparison, OCrystalNet only contains 2,365,441 parameters which is only two thirds of the parameters ALIGNN contains. Thus, OCrystalNet have advantages under limited experiments resources.


\begin{table}
  \caption{Comparison of MAEs of different models on JARVIS}
  \label{tab:JARVIS}
  \begin{tabular}{llllll}
    \hline
    Metric   & $E_g$ (OPT)  & $E_g$ (MBJ) & Formation Energy & Bulk Modulus & Shear Modulus  \\
    \hline
    CGCNN   & 0.20  & 0.41 & 0.063 & 14.47 & 11.75 \\
    ALIGNN  &0.14   & 0.31 & 0.033 & 10.40 & 9.48\\
    OCrystalNet & 0.14 & 0.32 & 0.044 & 9.03 & 7.95\\
    \hline
  \end{tabular}
\end{table}

From both the results on Material Project and JARVIS, OCrystalNet exhibits high performance on band gap prediction and bulk modulus prediction. Since band gap is the macroscopic representation of atomic valence orbital interaction in crystal material, we contribute the superiority to the rich information in atomic descriptor. As for bulk modulus, this property is highly correlated with the bonds between atoms and the topological structure of crystal graph, which is also captured by OCrystalNet. 

\subsection{Ablation Study}
\begin{figure}[htbp]
    \flushleft
    \includegraphics[width=1\textwidth]{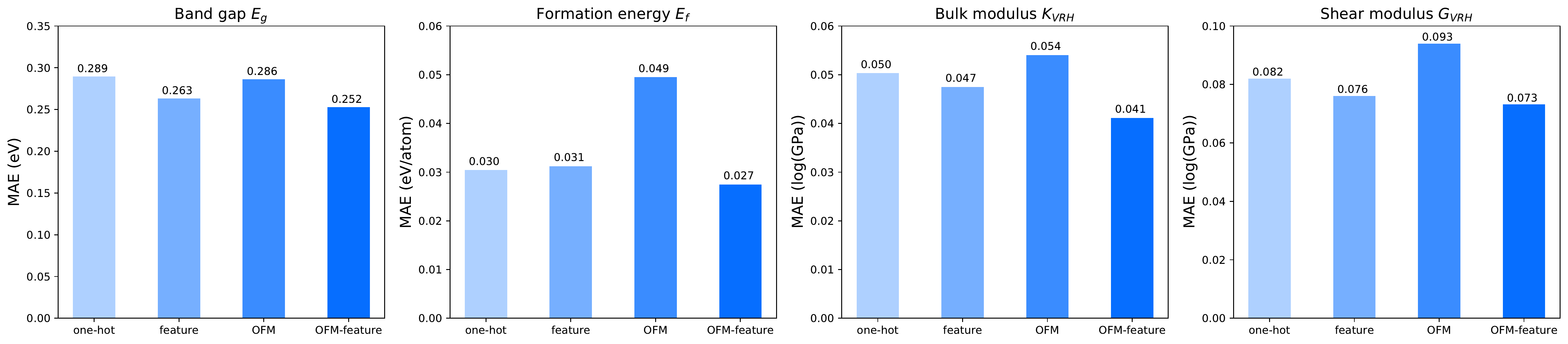}
    \caption{Comparison of different embedding method on OCrystalNet}
    \label{fg:ablation}
\end{figure}

We conducted ablation studies on four property prediction tasks to illustrate the effectiveness of embedding method proposed by this study. The result of Ablation study are shown in Table \ref{fg:ablation}. As shown in the result, atomic feature embedding alone is relatively effective when used alone in embedding process compared with one-hot embedding and OFM embedding. This is largely due to the fact that OFM descriptor alone can not effectively distinguish the difference between each atom. However, when we combine OFM and atomic feature together, the performance of OCrystalNet improved by a large margin. We contribute the improvement of OFM-physical descriptor to the additional valence orbital interaction information brought in by the OFM. 

\subsection{Case Study}

To further illustrate the effectiveness of OCrystalNet, select two crystal structures and give the predicted value and calculated value of specific property. As for band gap and formation energy, we choose $\rm Mn_2 O_7$. Compared with the computed value ($1.80$ eV) for band gap, OCrystalNet achieves acceptable result ($1.74$ ev) within DFT errors (~$0.6$ eV)\cite{jain2011high}. 

\begin{figure}[htbp]
    \centering

	\begin{minipage}{0.49\linewidth}
	    \centering
	    \includegraphics[width=1\textwidth]{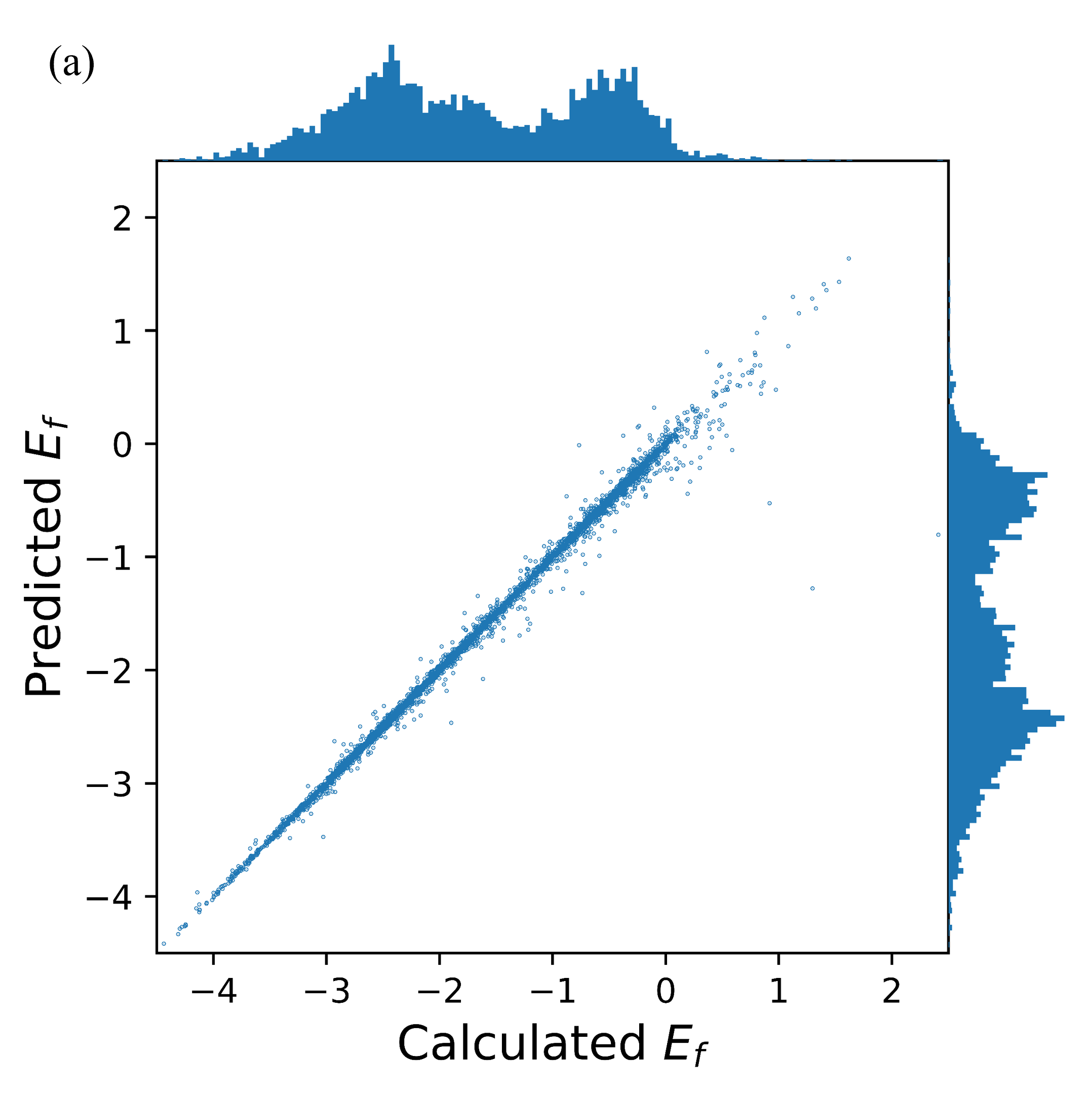}

	\end{minipage}
	\hfill
	\begin{minipage}{0.49\linewidth}
	    \centering
	    \includegraphics[width=1\textwidth]{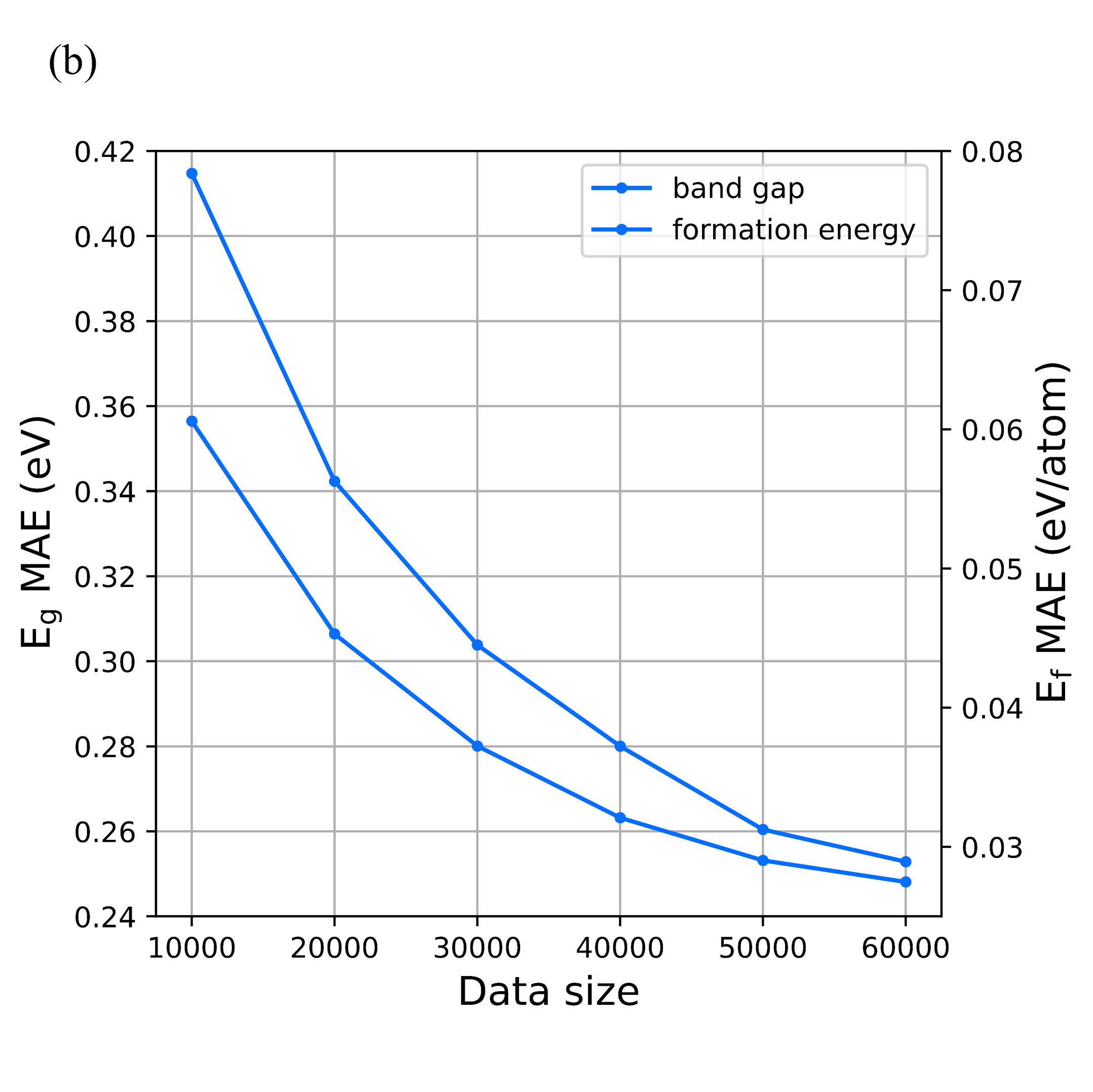}

	\end{minipage}
	\begin{minipage}{0.49\linewidth}
	    \centering
	    \includegraphics[width=1\textwidth]{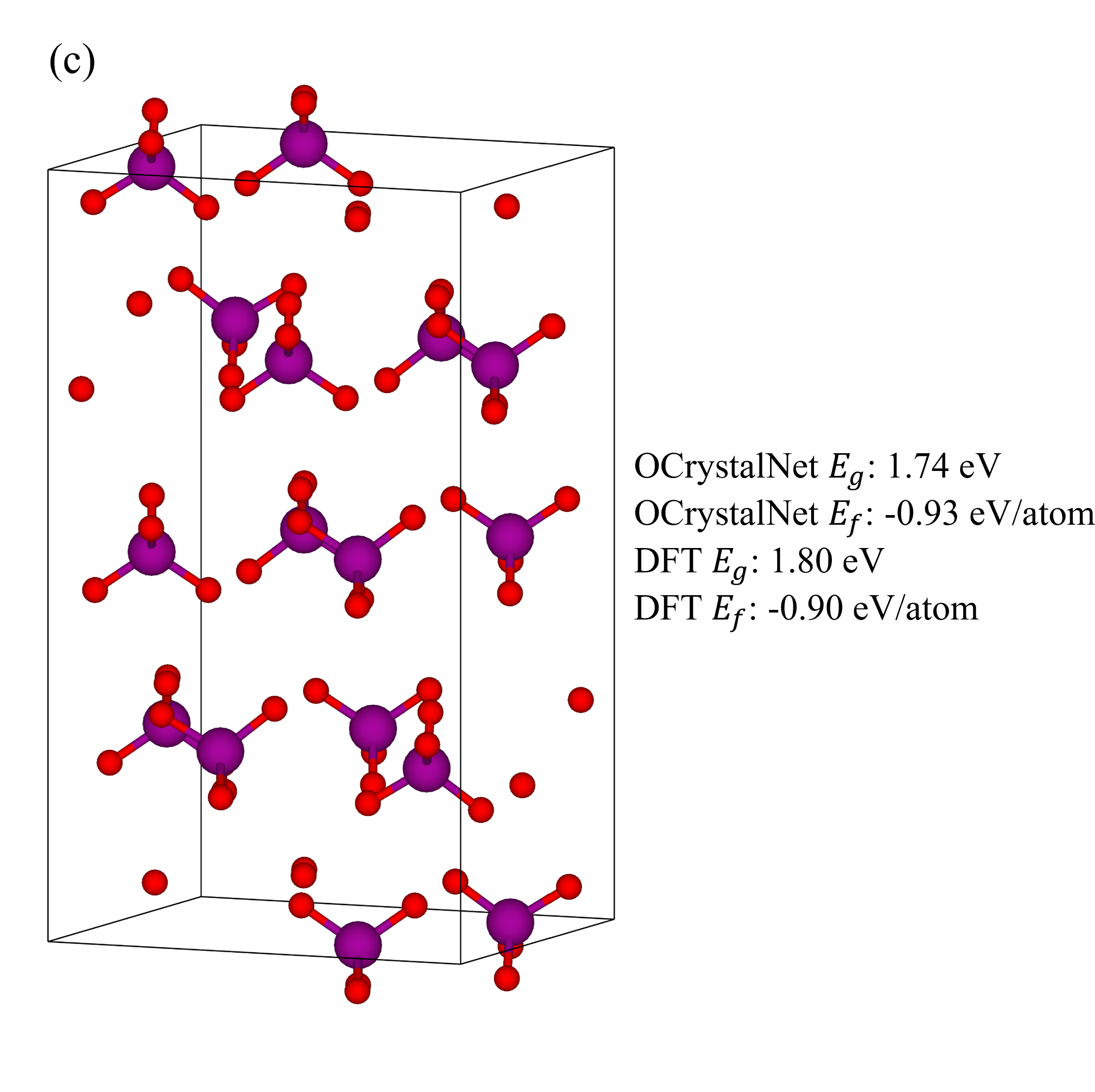}

	\end{minipage}
	\begin{minipage}{0.49\linewidth}
	    \centering
	    \includegraphics[width=1\textwidth]{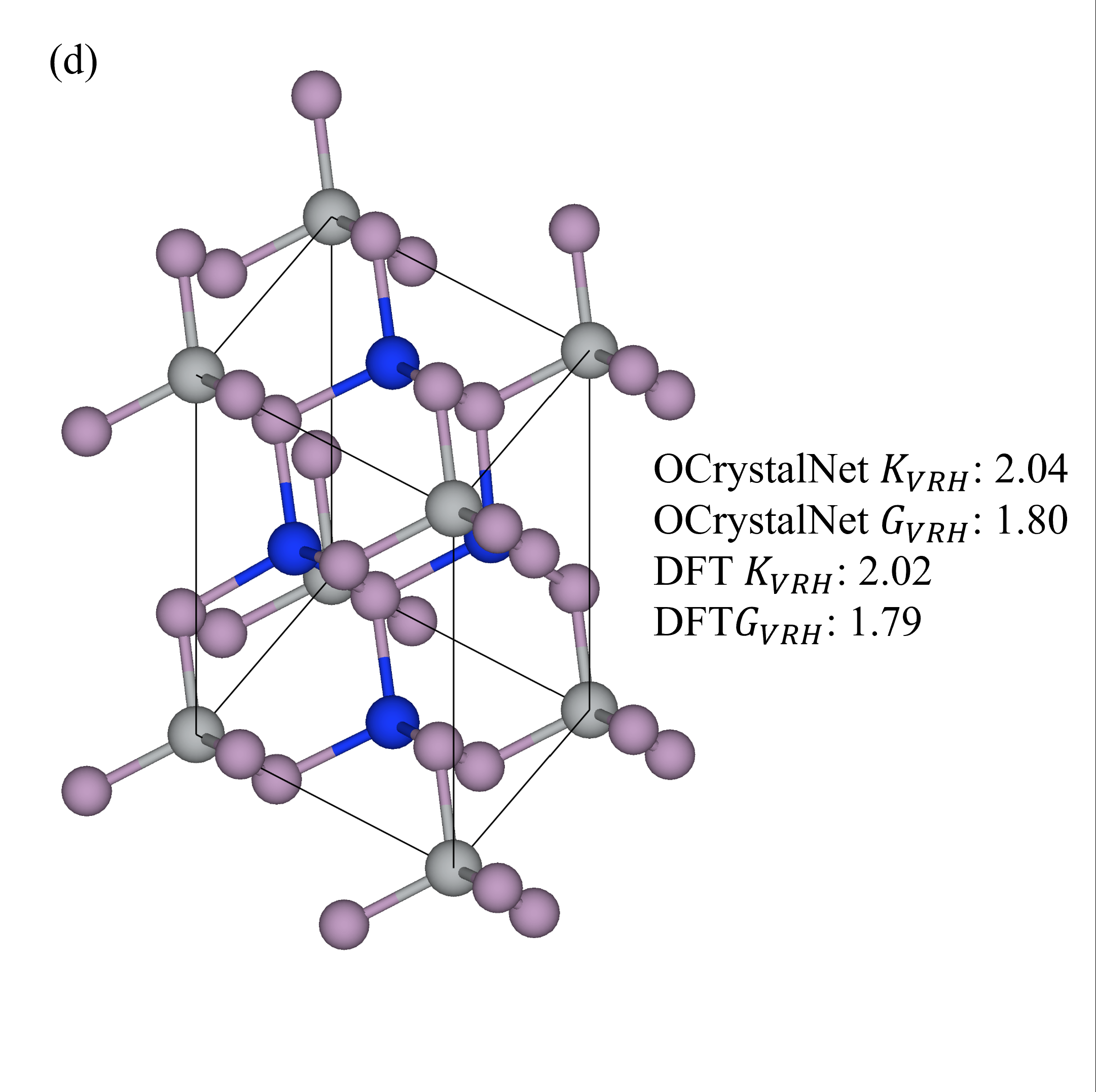}

	\end{minipage}
    \caption{Analysis of OCrystalNet on Material Project. (a) OCrystalNet predicted formation energy against DFT calculated formation energy. (b) MAE performance with different training data sizes. (c) Comparison of OcrystalNet and DFT on band gap and formation energy on crystal ${\rm Mn_2 O_7}$ (d) Comparison of OcrystalNet and DFT on Bulk Modulus and Shear Modulus on crystal $ \rm Si_3NiP_4$}
    \label{fg:summary}
\end{figure}

As for bulk modulus and shear modulus prediction, we choose a hypothetical material $\rm Si_3 NiP_4$. OCrystalNet also achieves highly accurate results (Figure \ref{fg:summary} (d)), in which both bulk modulus and shear modulus prediction are under $1\%$ error compared with DFT calculated results. Therefore, OCrystalNet is well capable of the task of predicting hypothetical materials' properties.

\subsection{Detailed analysis of OCrystalNet}

Detailed analysis of OCrystalNet is present with different perspectives of evaluation in this section. Figure \ref{fg:summary} (a) shows the predicted formation energy on test set (4620 crystals) on Material Project \cite{jain2013commentary} against DFT calculated formation energy. To further valid the effectiveness of OCrystalNet, we calculated the correlation coefficient, $R = 0.997$, between OCrystalNet predicted result and DFT calculated result.

Secondly, we trained OCrystalNet model on different size of the training set and test the result on Material Project \cite{jain2013commentary}. The results is shown in Figure \ref{fg:summary} (b). As the downtrend curve shows in Figure \ref{fg:summary} (b), the MAE of the predicted values decreases systematically as the the size of the training set increase. The lowest MAE achieves by OCrystalNet is $0.252$ at training set of 60000.



\section{Conclusion}
In This study, we proposed a deep learning model, OCrystalNet, under Message Passing Neural Network framework. We also proposed OFM-feature embeding leveraging valence orbital information and abundant atomic feature. Finally, we conducted extensive experiments on two open source datasets, namely Material Project and JARVIS. The experiments illustrate OCrystalNet's strong ability of learning vector representation of material graph and the superior generalization ability over other deep learning method. We also conducted ablation study to prove the effectiveness of the OFM-feature embedding proposed in the study. Our work achieves the best results in the tasks of predicting band gap, formation energy, bulk modulus and shear modulus and tightly combines the knowledge in chemistry of material with powerful deep learning model.

\bibliography{achemso-demo}

\end{document}